# Thermal Conductivity and Thermal Rectification in Graphene Nanoribbons: a Molecular Dynamics Study

*Jiuning Hu*[*,a,c], *Xiulin Ruan* [a,d] *and Yong P. Chen*[#,a,b,c]

[*]E-mail: hu49@purdue.edu; [#]E-mail: yongchen@purdue.edu.

[a]Birck Nanotechnology Center, Purdue University, West Lafayette, Indiana 47907, USA.

[b]Department of Physics, Purdue University, West Lafayette, Indiana 47907, USA.

[c]School of Electrical and Computer Engineering, Purdue University, West Lafayette, Indiana 47907, USA.

[d]School of Mechanical Engineering, Purdue University, West Lafayette, Indiana 47907, USA.

**ABSTRACT.** We have used molecular dynamics to calculate the thermal conductivity of symmetric and asymmetric graphene nanoribbons (GNRs) of several nanometers in size (up to ~4-nm wide and ~10-nm long). For symmetric nanoribbons, the calculated thermal conductivity (e.g. ~2000 W/m-K @400K for a 1.5 nm × 5.7 nm zigzag GNR) is on the similar order of magnitude of the experimentally measured value for graphene. We have investigated the effects of edge chirality and found that nanoribbons with zigzag edges have appreciably larger thermal conductivity than nanoribbons with armchair edges. For asymmetric nanoribbons, we have found significant thermal rectification. Among various triangularly-shaped GNRs we investigated, the GNR with armchair bottom edge and a vertex angle of 30° gives the maximal thermal rectification. We also studied the effect of defects and found that vacancies and edge roughness in the nanoribbons can significantly decrease the thermal conductivity. However, substantial thermal rectification is observed even in the presence of edge roughness.

Graphene, a monolayer of graphite and two dimensional honeycomb lattice of $sp^2$ bonded carbon,[1] is the building block of most carbon based nanomaterial, such as carbon nanotubes (CNTs) and buckyballs. Much attention has been given to the exceptional and unique electronic properties of graphene[2-5] uncovered in the past few years. Graphene nanoribbons (GNRs), narrow strips of graphene with few or few tens of nm in width, are particularly interesting and have been considered as important elements in future carbon-based nanoelectronics. For example, it has been shown that many of the electronic properties of GNRs may be tuned by its width or edge structures.[6-9]

In addition to its electronic properties, the thermal properties of graphene are also of fundamental and practical importance. Recent experiments[10,11] have demonstrated that graphene has a superior thermal conductivity, likely underlying the high thermal conductivity known in CNTs[12,13] and graphite[14] (ab-plane). This opens numerous possibilities of using graphene nanostructures for nanoscale thermal management. In this work, we have used molecular dynamics simulation to study thermal transport in few-nanometer-sized GNRs (typically 1-4 nm wide and 6-10 nm long). We have observed that the thermal conductivity of GNRs depends on the edge-chirality and can be affected by defects. We have also observed thermal rectification (TR) in asymmetric GNRs, where the thermal conductivity in one direction is significantly different from that in the opposite direction. Although TR has been experimentally observed in asymmetrically mass loaded nanotubes[15] and theoretically predicted in several other carbon nanostructures such as carbon nanohorns[16] and carbon nanotube intramolecular junctions[17], it has not been studied in any graphene systems to our knowledge. TR have potential applications in nanoscale thermal management such as on-chip cooling and energy conversion by controlling the heat transport, and is also fundamental in several recently proposed novel schemes of "thermal circuits" or information processing using phonons.[18-21] Our investigations of the thermal conductivity and controlling heat flow in graphene nanostructures can be important for the development of energy-efficient nanoelectronics based on graphene.

In our work, we have used classical molecular dynamics (MD) simulation based on the Brenner potential[22] of carbon-carbon interaction. We place atoms at the two ends of a GNR in the Nosé-Hoover thermostats[23,24] with temperatures $T_L$ (left end) and $T_R$ (right end) respectively (the temperature difference is denoted as $\Delta T$), and calculate the resulting heat current $J$. The thermal conductivity $\kappa$ is calculated from the well-known Fourier's law $\kappa = Jd/(\Delta Twh)$ where $d$, $w$ and $h$ ($=0.335nm$) are the length, width and thickness of the GNR respectively. The equations of motion for atoms in either the left or right Nosé-Hoover thermostat are:

$$\frac{d}{dt}p_i = F_i - \gamma p_i; \frac{d}{dt}\gamma = \frac{1}{\tau^2}\left[\frac{T(t)}{T_0} - 1\right]; T(t) = \frac{2}{3Nk_B}\sum_i \frac{p_i^2}{2m} \quad (1)$$

where the subscript $i$ runs over all the atoms in the thermostat, $p_i$ is the momentum of the $i$-th atom, $F_i$ is the total force acting on the $i$-th atom, $\gamma$ and $\tau$ are the dynamic parameter and relaxation time of the thermostat, $T(t)$ is the instant temperature of the thermostat at time $t$, $T_0$ ($=T_L$ or $T_R$) is the set temperature of the thermostat, $N$ is the number of atoms in the thermostat, $k_B$ is the Boltzmann constant and $m$ is the mass of the carbon atom. These equations of motion are integrated by a 3rd-order prediction-correction method. The time step is 0.5 fs, and the simulation runs for $10^7$ time steps giving a total MD time of 5 ns. The relaxation time $\tau$ is set to be 1 ps (see Additional Note 3). Typically, $T(t)$ can stabilize around the set value $T_0$ after 0.5 ns. Time averaging of the temperature and heat current is performed from 2.5 ns to 5 ns. The heat current injected to the thermostat is given by $J = \sum_i (-\gamma p_i) p_i / m = -3\gamma Nk_B T(t)$.[17]

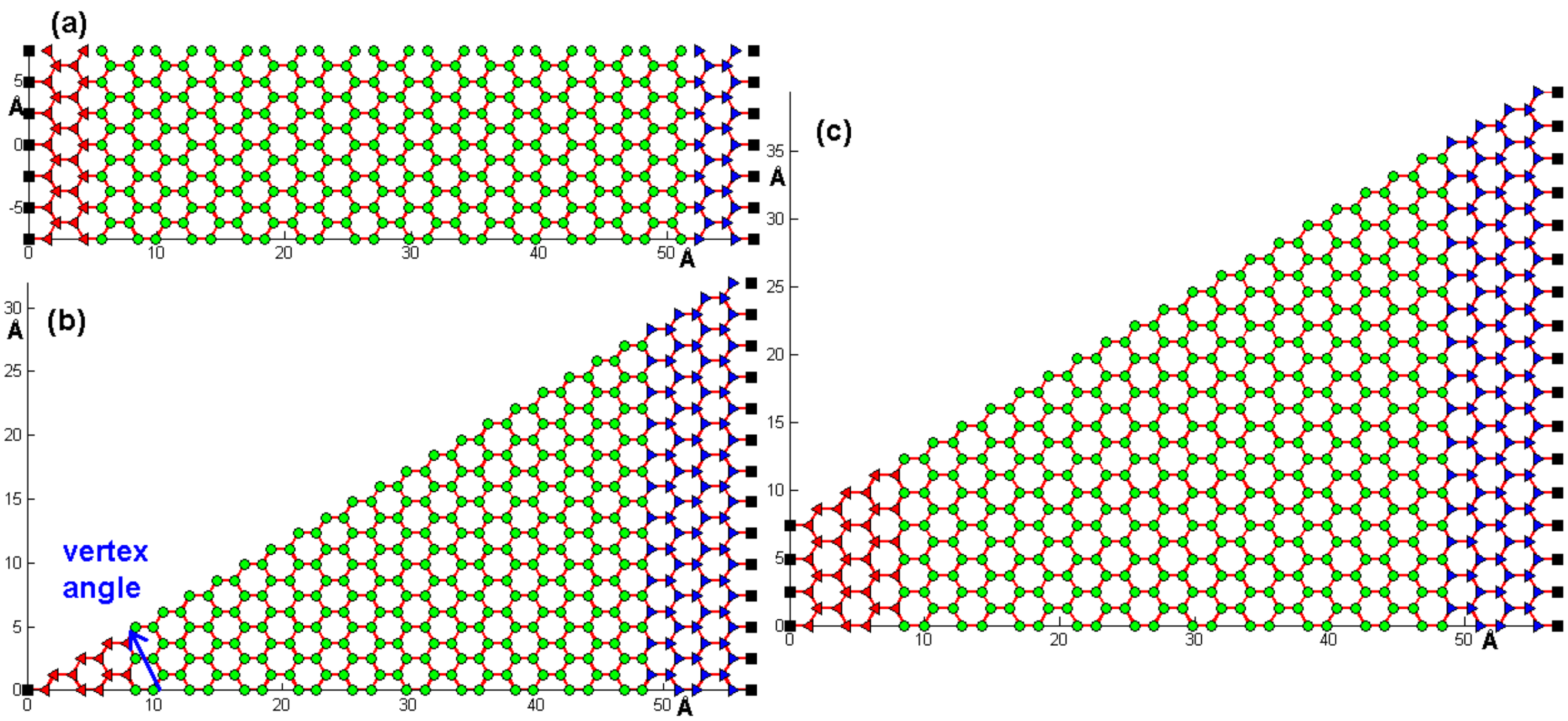

Figure 1. Structures of GNRs in this study: symmetric rectangle (a) and asymmetric triangle (b) and trapezoid (c).

In Figure 1, the atoms denoted with (black) squares at the ends are fixed to avoid the spurious global rotation of the GNRs in the simulation.[25] We have also performed the simulations with free and periodic boundary conditions, and found that the conclusions do not change qualitatively. The atoms denoted with triangles are placed in the Nosé-Hoover thermostats (obeying equation (1)) at temperature $T_0 = T_L$ (for the left-pointing triangles) and $T_0 = T_R$ (for the right-pointing triangles) respectively. The atoms denoted with circles obey the Newton's law of motion $\frac{d}{dt} p_i = F_i$. The average temperature of the two thermostats is $T = (T_L + T_R)/2$. We define $T_L = (1 \pm \alpha)T$ and $T_R = (1 \mp \alpha)T$ ($\alpha > 0$) so that the temperature difference is $\Delta T = |T_L - T_R| = 2\alpha T$. The parameter $\alpha = 10\%$ is used in all of the following calculations (see Additional Note 3). At steady state, there is no energy accumulation in GNRs, so in principle $J_L + J_R = 0$. However, due to the nature of the discretized MD simulation, $J_L + J_R$ slightly fluctuates around zero. Therefore, we define the heat current as $J = (J_L - J_R)/2$ and the error of the heat current as $\Delta J = |J_L + J_R|/2$ (corresponding to the error in the thermal conductivity $\Delta\kappa = \Delta Jd/(\Delta Twh)$). The temperature calculated from the classical MD ($T_{MD}$) has been corrected by taking into account the quantum effects of phonon occupation, using the scheme $T_{MD} = \frac{2T^3}{T_D^{\ 2}} \int_0^{T_D/T} \frac{x^2}{e^x - 1} dx$ , where $T_D$ (322 K) and $T$ are the Debye temperature and corrected temperature respectively (see Additional Note 3).

First, we calculate the thermal conductivity of symmetric GNRs of rectangular shapes. For a 5.7-nm long and 1.5-nm wide rectangular GNR (with zigzag long edge), the thermal conductivity is found to be around 2100 W/m-K at 400 K (Figure 2a), on the same order of magnitude with the experimental measured value (~3000-5000 W/m-K) of graphene.[10,11] On the other hand, the calculated thermal conductivity is nearly doubled after doubling the length of GNRs (see Additional Note 3). This suggests that our calculated thermal conductivity is limited by the finite length of GNRs and not corresponding to the value for graphene of macroscopic size. This is consistent with the phonon mean free path (MFP) in graphene extracted from the experiment (775 nm)[11] being much larger than the length (up to ~10 nm) of the GNRs simulated in this study. We have also found that the calculated thermal conductivity remains nearly the same after doubling the width of GNRs (with the length unchanged). As we will see for all of the GNRs in this study, the thermal conductivity always monotonically increases with temperature (*T*), in the range we studied (100-400 K). Similar behavior is predicted in a recent theory on the thermal conductivity of small graphene flakes.[26]

The effect of edge chirality on the thermal conductivity in rectangular GNRs is also investigated. The chirality of GNRs (see the right inset of Figure 2b for the definition of chirality angle) is defined according to the edge parallel to the long direction of the GNR (different from the convention for CNTs). In Figure 2a, thermal conductivity is plotted as a function of temperature for both zigzag (dashed line) and armchair (solid line) edged GNRs. We show that the thermal conductivity of zigzag GNR (dashed line) is 20-50% larger than that of the armchair GNR (solid line). Recent work by Jiang *et al.* obtained qualitatively similar results as ours using an approach based on ballistic phonon transport.[27] Similar effects have also been studied for CNTs, but there have been no generally accepted agreement on the preferred chirality of CNT for heat conduction.[28,29] CNTs are periodic in the azimuthal direction and they can be considered as GNRs with infinite width. All phonon modes can propagate in CNTs. However, in GNRs with finite width the phonon modes propagating along the heat flow direction dominate. We speculate that the difference between the thermal conductivity of armchair and zigzag GNRs is mainly due to the different phonon scattering rates at the armchair and zigzag edges and the effect of the finite size of GNRs. Figure 2b shows the thermal conductivity as a function of the chirality angle at the temperature of 400 K. Two peaks of the thermal conductivity at chirality angles of 0° (corresponding to armchair edge) and 30° (zigzag edge) can be clearly identified. The peak at 30° is higher than that at 0° as seen in Figure 2a. We only need to study the chirality angle from 0° to 60° due to the 6-fold rotational symmetry of graphene. The GNRs with chirality angle not equal to integer multiples of 30° have irregular edges. Phonons can be strongly scattered by these irregular edges, likely resulting in the relatively low thermal conductivity at these angles (thus the two peaks at 0° and 30°) as seen in Figure 2b.

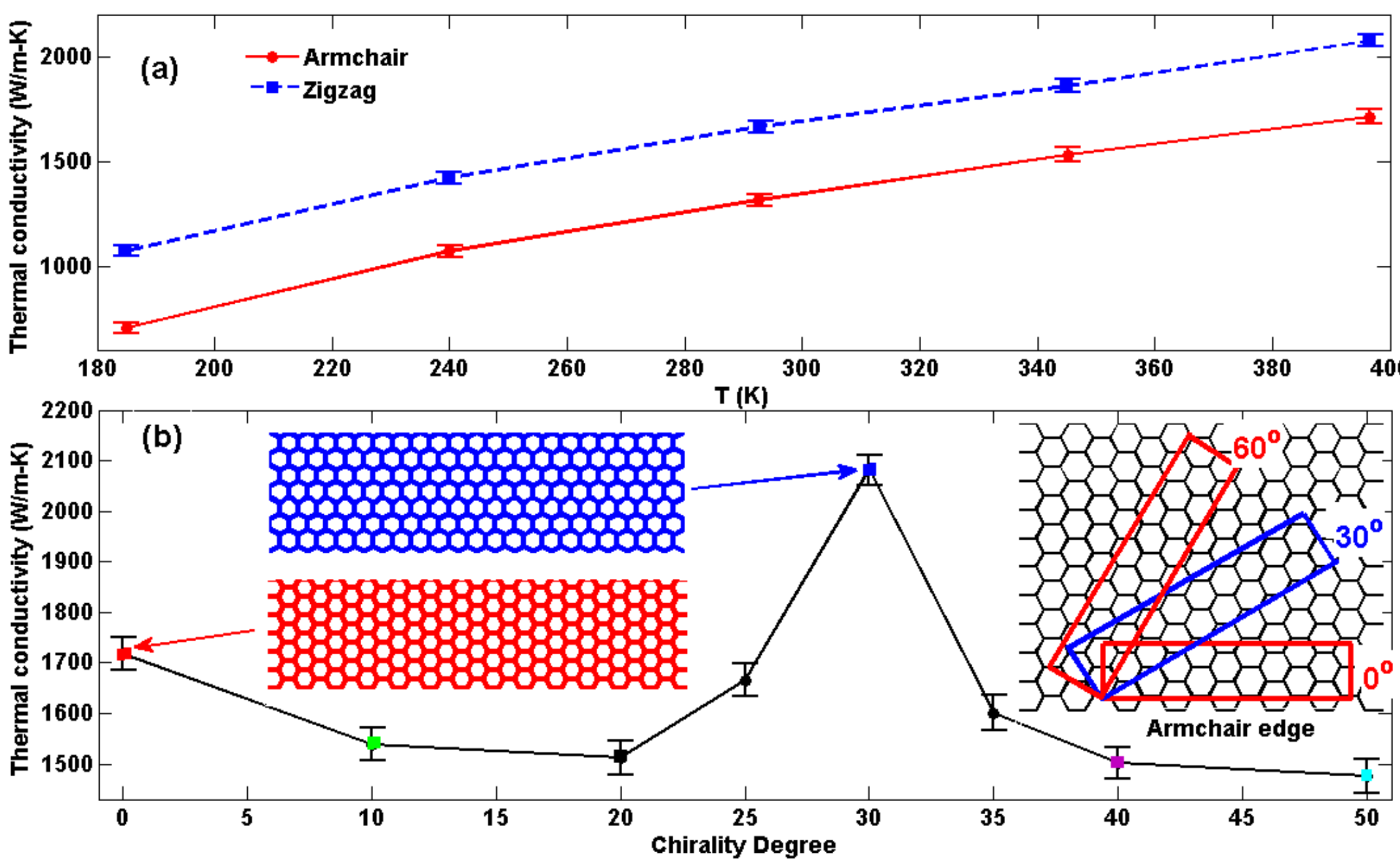

Figure 2. (a) Thermal conductivity of armchair and zigzag GNRs. (b) Thermal conductivity as a function of chirality angle (defined in right inset). The left two insets are two typical GNRs with edge chirality of zigzag (top) and armchair (bottom).

We have studied thermal conductivity of the asymmetric GNRs of triangular shapes (Figure 1b) and found significant thermal rectification. It is shown in Figure 3a that the thermal conductivity from the narrower (N) to the wider (W) end ($\kappa_{N\to W}$) of such a triangular GNR is less than that from the wider to the narrower end ($\kappa_{W\to N}$). The thermal rectification factor, defined as $\eta \equiv (\kappa_{W\to N} - \kappa_{N\to W})/\kappa_{N\to W}$ is as large as 120% (at $T$=180 K), as shown in the right inset of Figure 3a. We have also found similar thermal rectification in trapezoid-shaped GNRs (see Additional Note 4), but with smaller rectification factor. In contrary, in the symmetric rectangle GNRs there is no thermal rectification, i.e., the thermal conductivity form the left (L) to the right (R) is the same as that from the right to the left (dashed lines in Figure 3b) within the MD uncertainty. In calculating the thermal conductivity of asymmetric GNRs, the width ($w$) is taken as the width at the middle of GNRs. The temperature dependence of the thermal conductivity of the rectangular or trapezoidal GNRs is found to be similar to that of the rectangular GNRs. The overall thermal conductivity of asymmetric GNRs is lower than that of comparably sized symmetric GNRs.

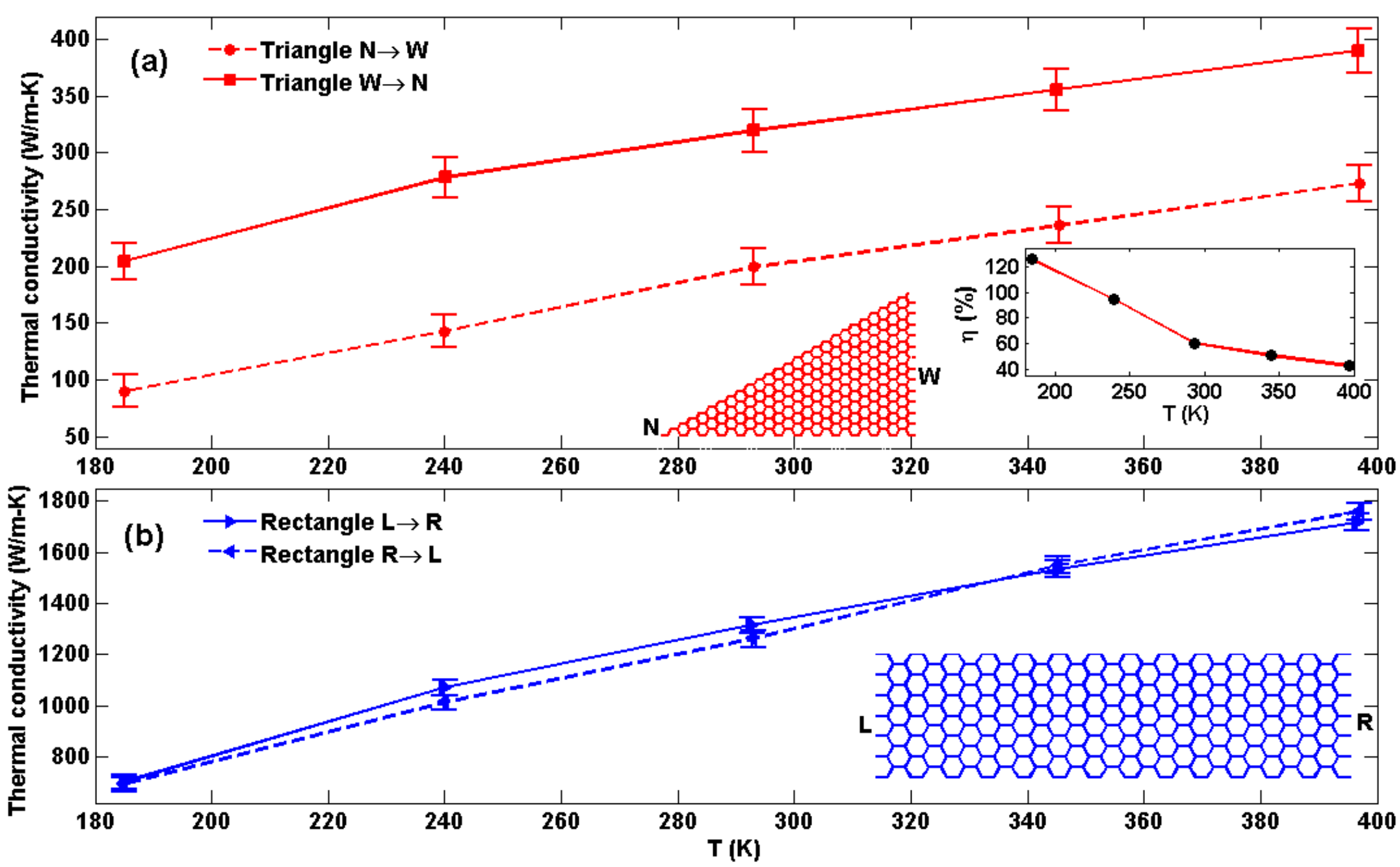

Figure 3. Thermal conductivity of (a) triangular asymmetric and (b) rectangular symmetric GNRs. The right inset of (a) shows the thermal rectification factor $\eta$ as a function of temperature.

Previous studies of thermal rectifications (TR)[15-18] have suggested that TR originates from the interplay between structural gradient and lattice nonlinearity. A similar mechanism can underlie the TR in our asymmetric GNRs. We have investigated several geometric variations of the triangular GNRs to search for the structure with largest thermal rectification. To make better comparison, the solid lines in Figure 4a and 4b are the calculated thermal conductivity of asymmetric triangular GNR shown in Figure 3a. This particular structure has an armchair bottom edge and a 30° vertex angle (defined in Figure 1b). Figure 4a shows the thermal rectification of the GNRs with different vertex angles but the same bottom edge length and chirality (armchair). The dashed lines (dotted lines) are for the GNRs with vertex angle of 45° (60°). Compared with the GNR with vertex angle of 30° (solid line), the GNR with vertex angle of 45° (60°) has lower (higher) thermal conductivity in both directions, but both GNRs with vertex angles of 45° and 60° exhibit less thermal rectification (see the inset of Figure 4a). For the GNR with vertex angle of 45°, its hypotenuse has irregular edge and the phonon scattering at this edge likely decreases both the thermal conductivity and the thermal rectification. In triangular GNRs with armchair bottom edge, only the GNRs with vertex angle of 30° and 60° do not have irregular hypotenuse edges. At large vertex angles, the triangular GNRs approach the symmetric rectangular GNRs, which has zero thermal rectification. Since both irregular edge scattering and large vertex angles can decrease the thermal rectification, we suggest that a vertex angle of 30° may be optimal for thermal rectification. We have demonstrated in Figure 2 that the symmetric zigzag GNRs have larger thermal conductivity than that of the armchair GNRs. This is also true for asymmetric triangular GNRs. In Figure 4b (4c), the GNR with armchair bottom edge and vertex angle of 30° (60°) (solid lines) has smaller thermal conductivity than the GNR with zigzag bottom edge and vertex angle of 30° (60°) (dashed lines), but the former has larger thermal rectification. Overall, among various triangular GNRs we investigated, the one with armchair bottom edge and vertex angle of 30° has the largest thermal rectification.

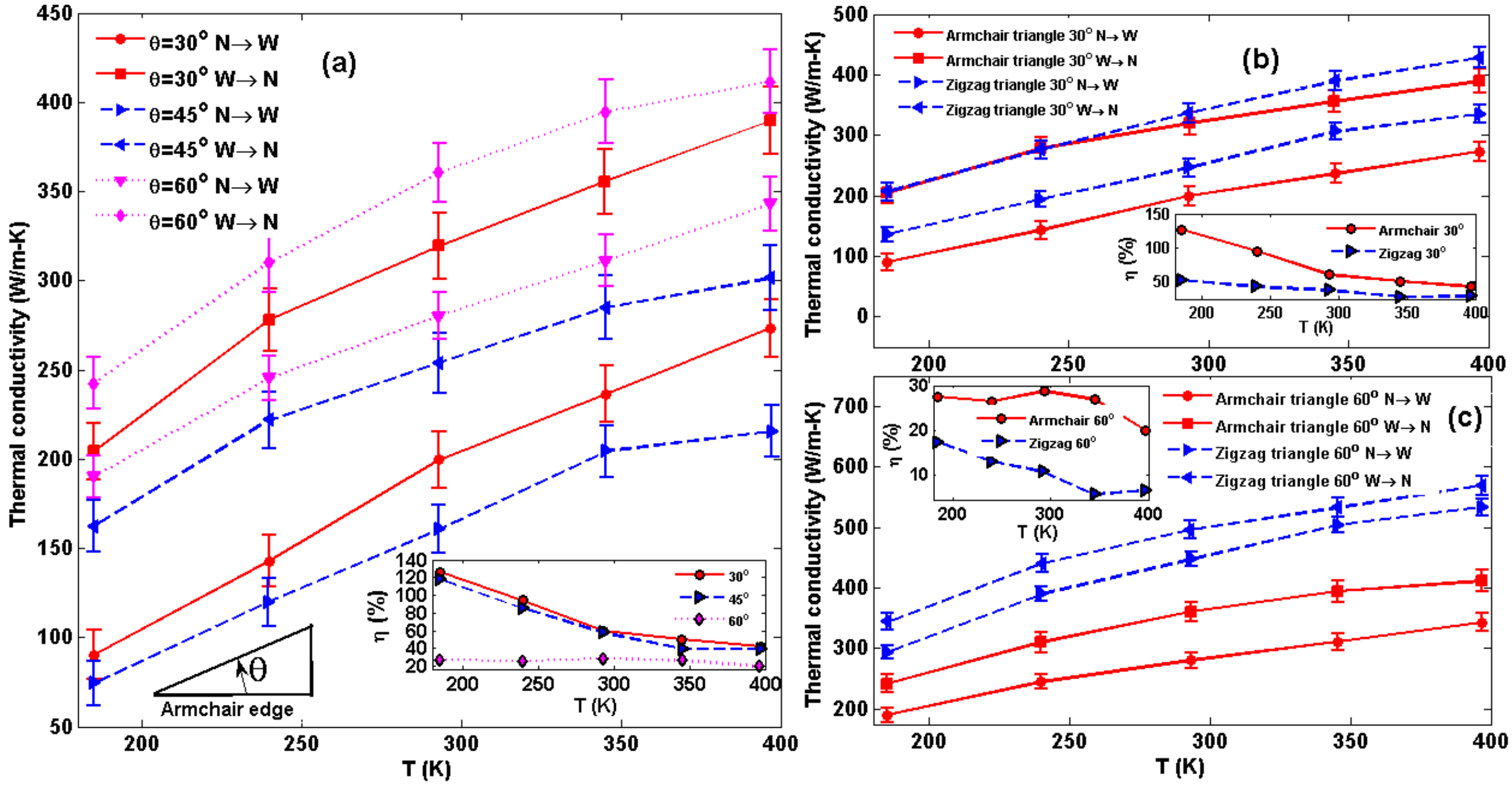

Figure 4. Thermal conductivity of triangular GNRs: (a) dependence of thermal conductivity on vertex angles for triangular GNRs with armchair bottom edges and effect of edge chirality of the bottom edge on thermal conductivity at vertex angle of (b) 30° and (c) 60°. The thermal rectification factors versus temperature for the various structures simulated are shown in insets.

In reality, GNRs inevitably have defects. We have studied the effect on the thermal conductivity of GNRs due to two types of defects: circular vacancies and edge roughness. In Figure 5a, we show the thermal conductivity of GNR with single (dotted line) and double (dashed line) circular vacancies and rough edges (dash-dot line). Here a single circular vacancy is created by removing all six carbon atoms of a hexagon. Compared to the perfect rectangular GNR (solid line), the thermal conductivity decreases after introducing circular vacancies. The edge roughness of the symmetric GNR can also decrease its thermal conductivity. This is in qualitative agreement with recent theoretical prediction by Nika *et al.* that the thermal conductivity of graphene flakes depends on the edge roughness and defect concentration, especially for small flakes.[30] The effect of edge roughness in the triangular GNR is also studied. We remove six atoms at the bottom edge and the hypotenuse of the triangular GNR of Figure 1b. The solid lines in Figure 5b represent the thermal conductivity of the perfect triangular GNR with armchair bottom edge and vertex angle of 30°. The edge roughness decreases both the thermal conductivity and the thermal rectification, which is nonetheless still as large as 80% at *T*~180 K.

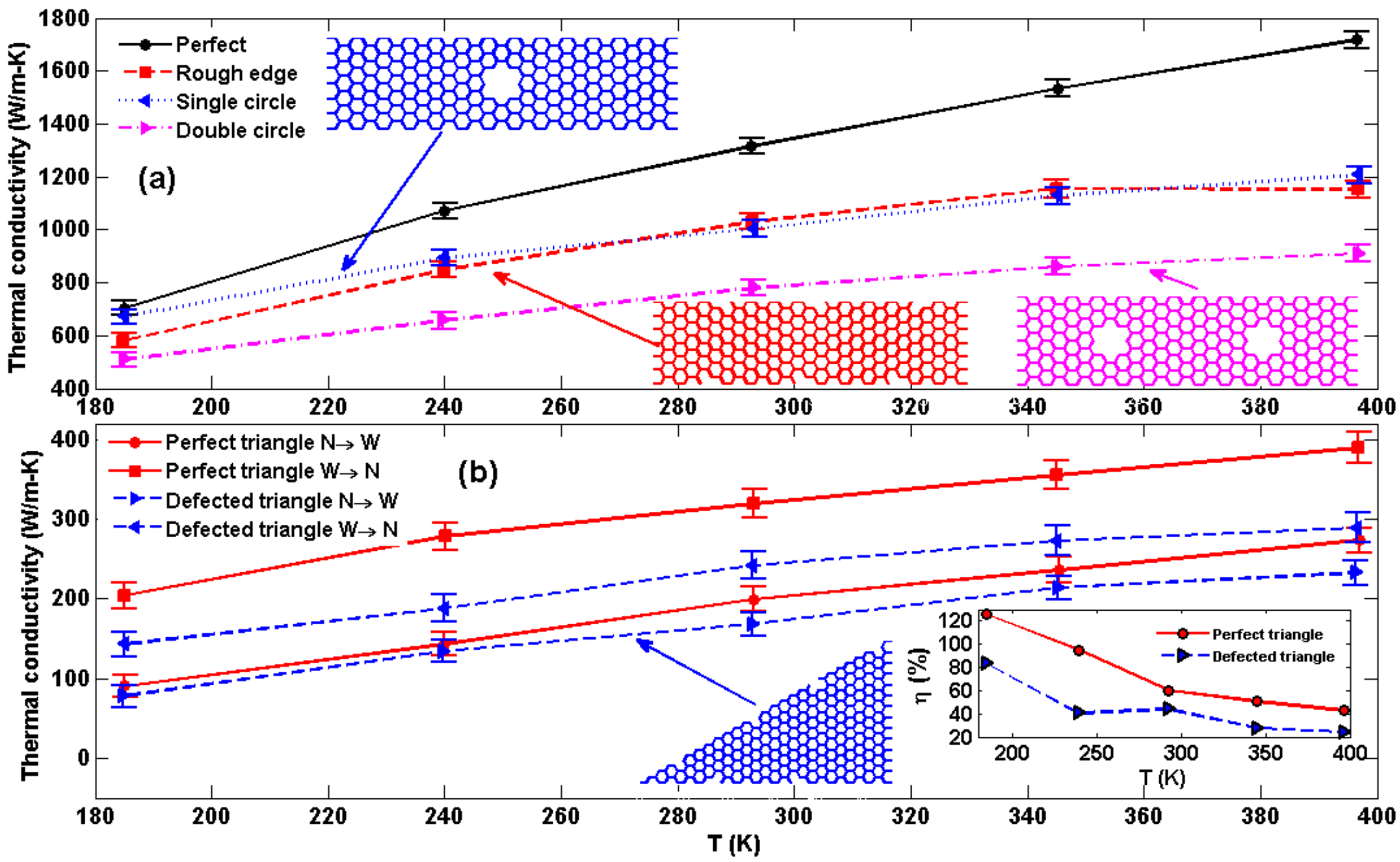

Figure 5. Effect of various defects on thermal conductivity of GNRs: (a) single and double vacancies and edge roughness for a symmetric rectangular GNR and (b) edge roughness for an asymmetric triangular GNR (the left inset shows the thermal rectification factor).

In summary, we have used classical molecular dynamics to calculate the thermal conductivity of both symmetric and asymmetric GNRs of few nanometers in size. The calculated thermal conductivity of symmetric rectangular GNRs is on the same order of magnitude as the value expected for graphene, but with differences likely caused by the finite sizes of GNRs. The thermal conductivity is shown to depend on edge chirality and the zigzag edge GNRs is shown to have larger thermal conductivity than that of the armchair edge GNRs. We have demonstrated the thermal rectification effect in asymmetric triangular and trapezoidal GNRs. The triangular GNR with vertex angle of 30° and armchair bottom edge is found to have the largest thermal rectification among all GNRs studied. The defects can reduce the thermal conductivity of GNRs as well as the thermal rectification in asymmetric GNRs.

**ADDITIONAL NOTES:**

1) After this work was published,[32] we became aware of similar results on thermal rectification in asymmetric GNRs obtained by Yang *et al*.[33]

2) The Fortran code used in this simulation is downloaded from http://www.eng.fsu.edu/~dommelen/research/nano/brenner/. The bugs of this code have been fixed by Dr. Leon van Dommelen. We have introduced Nosé-Hoover thermostats into the codes and set the two thermostats at different temperatures to simulate the non-equilibrium thermal transport.

3) More details about the MD calculations and the size dependence of thermal conductivity can be found in Ref. 34.

4) More details about thermal rectification in trapezoid-shaped GNRs can be found in Ref. 35.

**ACKNOWLEDGMENT**. This work is supported by the Semiconductor Research Corporation (SRC) - Nanoelectronics Research Initiative (NRI) via Midwest Institute for Nanoelectronics Discovery (MIND). We thank Dr. Gang Wu (National University of Singapore) for helpful communications on the MD code and Mr. Yaohua Tan (Purdue University) and Dr. Zhigang Jiang (Georgia Institute of Technology) for insightful discussions on graphene.

**REFERENCES:**

(1) Geim, A. K.; Novoselov, K. S. Nature Mater. 2007, 6, 183.

(2) Novoselov, K. S.; Geim, A. K.; Morozov, S. V.; Jiang, D.; Zhang, Y.; Dubonos, S. V.; Gregorieva, I. V.; Firsov, A. A. Science. 2004, 306, 666.

(3) Novoselov, K. S.; Geim, A. K.; Morozov, S. V.; Jiang, D.; Katsnelson, M. I.; Grigorieva, I. V.; Dubonos, S. V.; Firsov, A. A. Nature. 2005, 438, 197.

(4) Zhang, Y.; Tan, Y. W.; Stormer, H. L.; Kim, P. Nature. 2005, 438, 201.

(5) Castro Neto, A. H.; Guinea, F.; Peres, N. M. R.; Novoselov, K. S.; Geim, A. K. Rev. Mod. Phys. 2009, 81, 109.

(6) Nakada, K.; Fujita, M.; Dresselhaus, G.; Dresselhaus, M. S. Phys. Rev. B 1996, 54, 17954.

(7) Son, Y. W.; Cohen, M. L.; Louie, S. G. Nature, 2006, 444, 347.

(8) Han., M. Y.; Özyilmaz, B.; Zhang, Y.; Kim, P. Phys. Rev. Lett. 2007, 98, 206805.

(9) Chen, Z.; Lin, Y. M.; Rooks, M. J.; Avouris, P. Physica E, 2007, 40, 228.

(10) Balandin, A. A.; Ghosh, S.; Bao, W.;Calizo, I.; Teweldebrhan, D.; Miao, F.; Lau, C. N. Nano Lett. 2008, 8, 902.

(11) Ghosh, S.; Calizo, I.; Teweldebrhan, D.; Pokatilov, E. P.; Nika, D. L.; Balandin, A. A.; Bao, W.; Miao, F.; Lau, C. N. Appl. Phys. Lett. 2008, 92, 151911.

(12) Kim, P.; Shi, L.; Majumdar, A.; McEuen, P. L. Phys. Rev. Lett. 2001, 87, 215502.

(13) Pop, E.; Mann, D.; Wang, Q.; Goodson, K.; Dai, H. Nano Lett. 2006, 6, 96.

(14) Pierson, H. O. Handbook of Carbon, Graphite, Diamonds and Fullerenes: Processing, Properties and Applications; Noyes Publications: USA, 1995.

(15) Chang, C. W.; Okawa, D.; Majumdar, A.; Zettl, A. Science 2006, 314, 1121.

(16) Wu, G.; Li, B. J. Phys.: Condens. Matter 2008, 20, 175211.

(17) Wu, G.; Li, B. Phys. Rev. B 2007, 76, 085424.

(18) Li, B.; Wang, L.; Casati, G. Phys. Rev. Lett. 2004, 93, 184301.

(19) Li, B.; Wang, L.; Casati, G. Appl. Phys. Lett. 2006, 88, 143501.

(20) Wang, L.; Li, B. Phys. Rev. Lett. 2007, 99, 177208.

(21) Wang, L.; Li, B. Phys. Rev. Lett. 2008, 101, 267203.

(22) Brenner, D. W. Phys. Rev. B 1990, 42, 9458.

(23) Nosé, S. J. Chem. Phys. 1984, 81, 511.

(24) Hoover, W. G. Phys. Rev. A 1985, 31, 1695.

(25) Hunenberger, P. H. Adv. Polym. Sci. 2005, 173, 105.

(26) Nika, D. L.; Glosh, S.; Pokatilov, E. P.; Balandin, A. A. Preprint at http://arxiv.org/abs/0904.0607, 2009.

(27) Jiang, J.; Wang, J.; Li, B. Preprint at http://arxiv.org/abs/0902.1836, 2009.

(28) Zhang, G.; Li, B. J. Chem. Phys. 2005, 123, 114714.

(29) Yamamoto, T.; Watanabe, S.; Watanabe, K. Phys. Rev. Lett. 2004, 92, 075502.

(30) Nika, D. L.; Pokatilov, E. P.; Askerov, A. S.; Balandin, A. A. Phy. Rev. B 2009, 79, 155413.

(31) Falkovsky, L.A. Phys. Lett. A 2008, 372, 5189.

(32) Hu, J.; Ruan, X.; Chen, Y. P. Nano Lett. 2009, 9, 2730.

(33) Yang, N.; Zhang, G.; Li, B. Appl. Phy. Lett. 2009, 95, 033107.

(34) Hu, J.; Ruan, X.; Jiang, Z.; Chen, Y. P. Int. Conf. on Frontiers of Characterization and Metrology for Nanoelectronics 2009, AIP Conf. Proc. 2009, 1173, 135.

(35) Hu, J.; Ruan, X.; Chen, Y. P. 17$^{th}$ Symposium on Thermal Physical Properties, Preprint available at http://www.physics.purdue.edu/quantum/files/2009.Boulder.IJT_.v3.pdf, 2009.